\documentclass[journal=jacsat,manuscript=article]{achemso}

\usepackage{geometry}
	\usepackage{graphicx}
	\usepackage{xcolor}
	\usepackage[version=4]{mhchem} 

\usepackage{graphicx}
\usepackage{hyperref}
\usepackage{indentfirst}
\hypersetup{colorlinks}
\newcommand{\orcidicon}[1]{\href{https://orcid.org/#1}{\includegraphics[height=\fontcharht\font`\B]{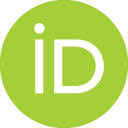}}}

\author{Mohammad Babar\,\orcidicon{0000-0001-6779-3859}}
\affiliation{Mechanical Engineering Department, Carnegie Mellon University, Pittsburgh, Pennsylvania 15213, USA}
\author{Hasnain Hafiz\,\orcidicon{0000-0002-0202-794X}}
\affiliation{Department of Physics, Northeastern University, Boston, MA, USA}
\author{Zeeshan Ahmad\,\orcidicon{0000-0001-9758-8952}}
\affiliation{Pritzker School of Molecular Engineering, University of Chicago, Chicago, Illinois 60637, USA}
\author{Bernardo Barbiellini\,\orcidicon{0000-0002-3309-1362}}
\affiliation{Department of Physics, School of Engineering Science, LUT University, Lappeenranta, Finland}
\alsoaffiliation{Department of Physics, Northeastern University, Boston, MA, USA}
\author{Arun Bansil}
\affiliation{Department of Physics, Northeastern University, Boston, MA, USA}
\author{Venkatasubramanian Viswanathan\,\orcidicon{0000-0003-1060-5495}}
\affiliation{Mechanical Engineering Department, Carnegie Mellon University, Pittsburgh, Pennsylvania 15213, USA}
\email{venkvis@cmu.edu}

\title[An \textsf{achemso} demo]
  {Effect of disorder and doping on electronic structure and diffusion properties of \ce{Li3V2O5}}

\abbreviations{IR,NMR,UV}

\begin{document}

\begin{abstract}
  \ce{V2O5} in its $\omega$ phase (\ce{Li3V2O5}) with excess lithium is a potential alternative to the graphite anode for lithium-ion batteries at low temperature and fast charging conditions due to its safer voltage (0.6 V vs \ce{Li$^{+}$/Li(s)}) and high lithium transport rate. In-operando cationic disorder, as observed in most ordered materials, can produce significant changes in charge compensation mechanisms, anionic activity, lithium diffusion and operational voltages. In this work, we report the variation in structural distortion, electronic structure and migration barrier accompanied by disorder using first-principles calculations. Due to segregation of lithium atoms in the disordered state, we observe greater distortion, emergence of metallic behaviour and potential anionic activity from non-bonding oxygen states near the Fermi level. Redox capacity can be tuned by doping with 3d metals which can adjust the participating cationic states, and by fluorine substitution which can stabilize or suppress anionic states. Moreover, suppression of anionic activity is found to decrease structural distortion, crucial for mitigating voltage fade and hysteresis. Diffusion barrier calculations in the presence of disorder indicate the activation of the remaining 3D-paths for lithium hopping which are unavailable in the ordered configuration, explaining its fast-charging ability observed in experiments.

\end{abstract}

\section{Introduction}

High capacity and long cycle life of lithium-ion batteries form the dominant characteristics for both near-term and long-term technologies in consumer electric vehicles, aviation, aerospace and energy storage devices \cite{choi2016promise, zeng2019commercialization, zeng2018stabilization}. Transition metal oxides (TMOs) with disordered rock salt (DRS) structure e.g. \ce{Li2W2O7}, \ce{Li2MoO4}, \ce{Li2NbO4} and \ce{Li2VO2F} have been a popular choice for high-capacity lithium-rich positive electrode (cathodes) materials  \cite{chen2015disordered, yahia2019unified, radin2019manganese, yabuuchi2015high}. Out of these, lithium vanadium pentaoxide (\ce{Li$_{x}$V2O5}) has been under critical examination for more than 30 years since first proposed \cite{whittingham1976role}. Structurally significant phases in \ce{V2O5} have been accurately determined experimentally \cite{rocquefelte2003first} starting from the layered $\alpha, \epsilon, \delta$ and $\gamma$ phases to the fully reduced DRS $\omega$ phase for $x=3$ in \ce{Li$_{x}$V2O5}. As a cathode, the layered phases are reversible, have an acceptable voltage range (2-4 V), and undergo small volume expansion \cite{rocquefelte2003first, horrocks2013finite}. Nanostructure engineering of \ce{V2O5} to improve these properties has also shown significant enhancement in cathodic performance \cite{pomerantseva2012electrochemical, zhang2011carbon, wang2012electrospun, shepard2020ab}. More recently, however, \citet{liu2020disordered} showed that excess lithium insertion in \ce{V2O5} in its $\omega$ phase ($x=3$-$5$) is also reversible and can be used as a fast-charging anode. Its average operational voltage is 0.4 V higher than graphite (0.2 V vs \ce{Li$^{+}$/Li(s)}) i.e. a broader potential range for lithium insertion offering better stability and safety against plating under low temperature and fast charging conditions \cite{legrand2014physical, verbrugge1997effect, petzl2015lithium}. Consequently, the irreversible transformation from layered to DRS ($\gamma$-$\omega$) marks the fully lithiated (de-lithiated) phase as a cathode (anode). 

Starting with the ordered structure, operational temperature and voltage fluctuations during charge-discharge cycles can potentially cause in-operando cationic disorder \cite{abdellahi2016effect}. This is commonly accompanied by structural distortions and activation of alternate reaction centers like anionic redox, which can change electrode capacity and cause detrimental effects like hysteresis, voltage fade and sluggish kinetics \cite{assat2018fundamental}. Oxygen redox can happen at as low voltages as 3.0V in \ce{Li2O} \cite{clement2020cation}. Therefore, contrary to the norm that anionic activity is limited to cathodes, we might observe disorder mediated oxygen redox in lithium excess \ce{Li_{3+x}V2O5} anode, mainly during the initial stages of lithium insertion ($x << 1$). Electronic structure of pristine \ce{Li3V2O5} can highlight important differences in reaction mechanisms of disordered and ordered systems. Moreover, these differences can be engineered with doping of anions (e.g. fluorination) and cations (transition metal substitution) within the bulk. Anionic substitution by fluorine is known to increase cationic redox capacity \cite{clement2020cation, chen2015disordered}, while 3d metal substitution can modulate the activity of either redox center. There is also a greater possibility for easier microscopic diffusion in the disorder due to the formation of favorable cation clusters relevant to lithium transport \cite{lee2014unlocking, ji2019hidden} and smaller hindrance by distortion. Macroscopic diffusion paths are already activated by a connected percolation network of 0TM sites when lithium excess is above a threshold in the disorder \cite{urban2014configurational, ji2019hidden}.

Here, we present a first-principles comparison of electronic structure between the pristine ordered and disordered \ce{Li3V2O5} and its correlation with structural distortion, for better insight into the effects of cationic disorder. We obtain the reference (ordered and disordered) structures by sorting the configurations based on the formation energies and electron population in oxygen atoms as descriptors. We assume the reference structures to be representative of the ordered and disordered regimes in \ce{Li3V2O5}. Using normal coordinates \cite{urban2017electronic}, we transform displacement errors of the most disfigured octahedron into distortion modes (stretching, bending, twisting etc.), allowing better quantification of structural distortion induced by disorder and doping. Next, we employ projected density of states (pDOS) obtained from Density Functional Theory (DFT) calculations to analyze the nature of oxygen bonding, and their locations with respect to Fermi level and transition metal (TM) states, in ordered, disordered and doped structures. Potential peroxo bonds between oxygen atoms are examined via crystal orbital/Hamiltonian overlap population (COOP/COHP) calculations. A consistent correlation between structural distortion and the activity of anionic centers in the disordered structure is observed, which can be suppressed or enhanced with cationic (3d metal) or anionic (fluorine) doping. Finally, the possibility for improved microscopic diffusion due to weakening TM-O bonds in the disordered state is investigated using Nudged Elastic Band (NEB) calculations. This enables us to study the effect of disorder on the electronic structure and microscopic diffusion of \ce{Li3V2O5} within an ab-initio framework. Our study could be an important step towards exploiting the strengths of disordered systems while mitigating voltage fade and structural degradation.
 

\section{Methods}
All DFT calculations were performed using the Quantum Espresso \cite{giannozzi2009quantum} code with Ultrasoft pseudopotentials \cite{garrity2014pseudopotentials}. We employed Perdew–Burke–Ernzerhof (PBE) \cite{perdew1996generalized} generalized gradient approximation (GGA) exchange correlation functional with a Hubbard $U$ extension of 3.25 eV \cite{liu2020disordered, du2021tunable} on vanadium site for obtaining energies, atomic relaxations, electronic structure and migration barriers. It is well known that GGA+U and hybrid functional methods have a qualitatively equivalent behaviour in transition metal complexes (TMCs) \cite{gani2016does}. Previous first-principles studies \cite{du2021tunable, scanlon2008ab} demonstrated that GGA+$U$ can essentially capture accurate electronic structures in vanadate-based oxides. \citet{liu2020disordered} also reported that  GGA+$U$ approach could achieve qualitatively accurate open circuit voltage (OCV) trend, sub-phase transitions and volume changes in \ce{Li_{3+x}V2O5} matching with experiments.  

For other 3d metals, $U$ values fitted from binary formation enthalpies have been taken from \citet{jain2011high}. The kinetic energy cutoff for plane waves and charge density was fixed at 544 eV (40 Ry) and 5440 eV (400 Ry) respectively \cite{liu2020disordered}. All systems were initialized with high spin ferromagnetic configuration for the least error in relative energies \cite{ling2014phase}. The energies and forces were converged to 0.0013 eV ($10^{-4}$ Ryd) per cell and 0.0257 eV$/$Å ($10^{-3}$ Ry/Bohr)  respectively. All relaxations were performed using Broyden–Fletcher–Goldfarb–Shanno (BFGS) optimization \cite{fletcher1981practical}. Crystal orbital overlap populations (COOP) were obtained using the LOBSTER code \cite{maintz2016lobster} for an analysis of the O-O and V-O bond strengths, as performed earlier on other Li-TMOs \cite{saubanere2016intriguing}. A set of 80 symmetrically distinct orderings of the same stoichiometry (\ce{Li3V2O5}) were generated using enumlib wrapper \cite{hart2008algorithm} to compare between cationic disorder. 

Site occupancy of lithium atoms (tetrahedral or octahedral) was determined using Atomic Simulation Environment's (ASE) \cite{larsen2017atomic} Neighbourlist functionality and Brunner’s algorithm in pymatgen \cite{ong2013python}. Lithium atoms with four-fold coordination are classified as tetrahedral sites, and the 5-coordinated square-pyramidal and 6-coordinated octahedral environments as octahedral sites. For migration barriers, we used climbing image NEB with 5 transition steps. Supercells ($2\times2\times2$) of the single formula unit cell (\ce{Li3V2O5}) were chosen with approximately uniform and large enough ($\sim$9Å $\times$ 9Å $\times$9Å) dimensions in order to avoid interactions with periodic images. We considered three types of lithium migration mechanisms, direct tetrahedron-to-tetrahedron (t-t), intermediate octahedra knock-off at a corner site (t-o-t corner-sharing) or at opposite sites (t-o-t opposing) \cite{liu2020disordered}. The energies and forces were converged to 0.0013 eV ($10^{-4}$ Ryd) per cell and 0.0257 eV$/$Å ($10^{-3}$ Ryd$/$Bohr), respectively. 

For quantification of structural distortion \cite{urban2017electronic}, the displacement error between an ideal octahedron (\textit{O}) and a disordered one (\textit{D}) were transformed into normal modes as, 
\begin{equation}
    D = O + \Sigma_{i} c_{i}\Tilde{Q}_{i}
\end{equation}
where $c_{i}$ are mode coefficients and $\Tilde{Q}_{i}$ are normalized normal coordinates. The magnitude of coefficients are indicative of the amount of distortion in a given mode. To get the displacement vectors ($D-O$), $O$ is transformed with a 3D rotation matrix (yaw, pitch and roll), which minimizes the displacement error under BFGS optimization. The ideal octahedron is assumed to have a uniform cubic parameter, experimentally observed from DRS-\ce{Li3V2O5} (4.095 Å) \cite{liu2020disordered}.  

\section{Results and discussion}

\subsubsection{Disorder formation energies and electron population in oxygen atoms}

\begin{figure*}[hbt!]
\centering
\includegraphics[width=4.6in]{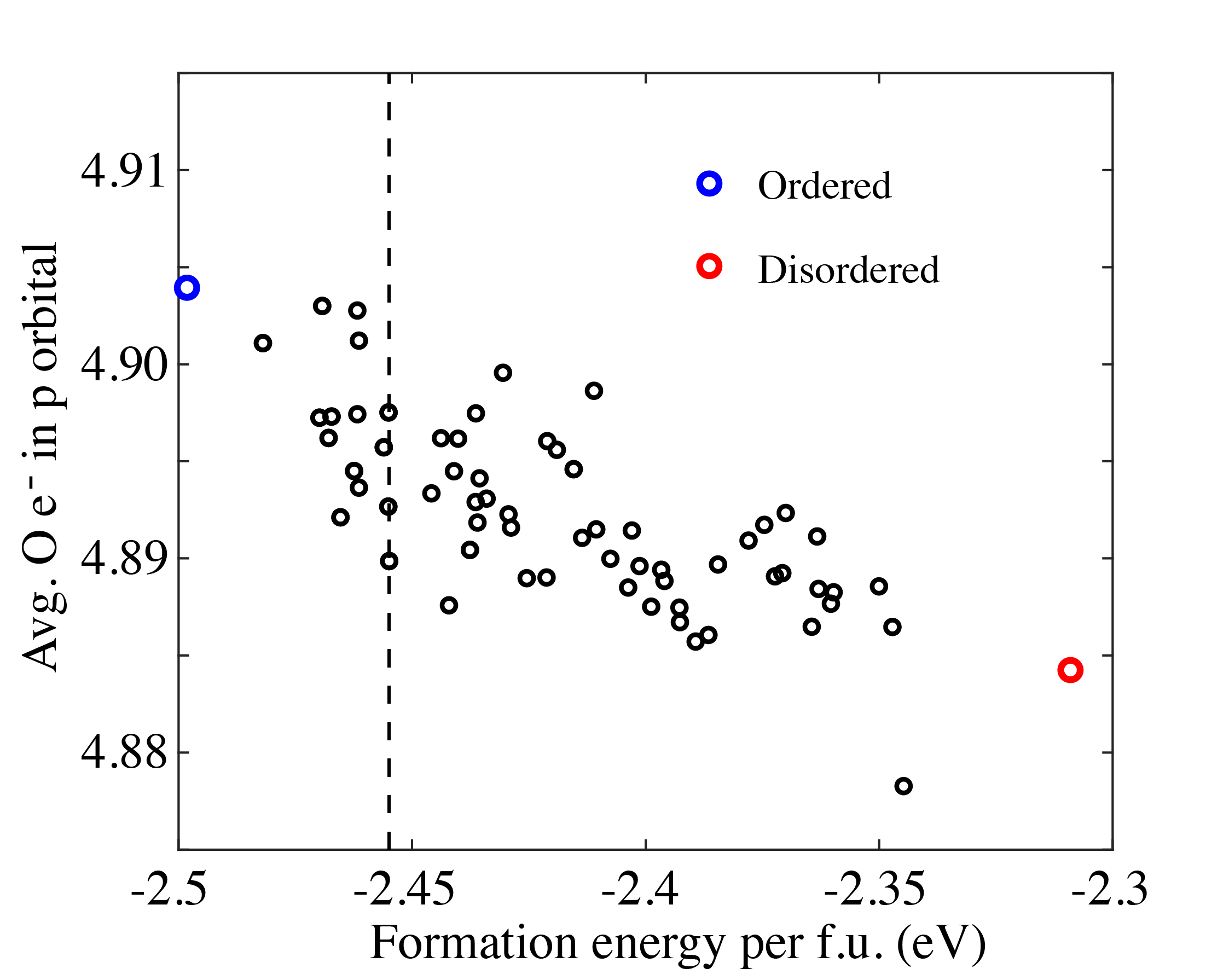}
\caption{\label{fig: fE_ox} Formation energy per formula unit of the disordered configurations as a function of average population of electrons in the oxygen \textit{p}-orbitals. Oxygen atoms digress from the full octet (\ce{O^{2-}}) state with increasing extent of disorder and structural distortion. Stability of the configurations fall almost linearly with the average occupation of oxygen states. The edge points on either sides, i.e. ordered (blue) and disordered (red) structures are marked respectively, to be used as references for comparison. Structures on the left of the dashed line are accessible via configurational entropy at room temperature.}
\end{figure*}

\begin{figure*}[hbt!]
\centering
\includegraphics[width=6.9in]{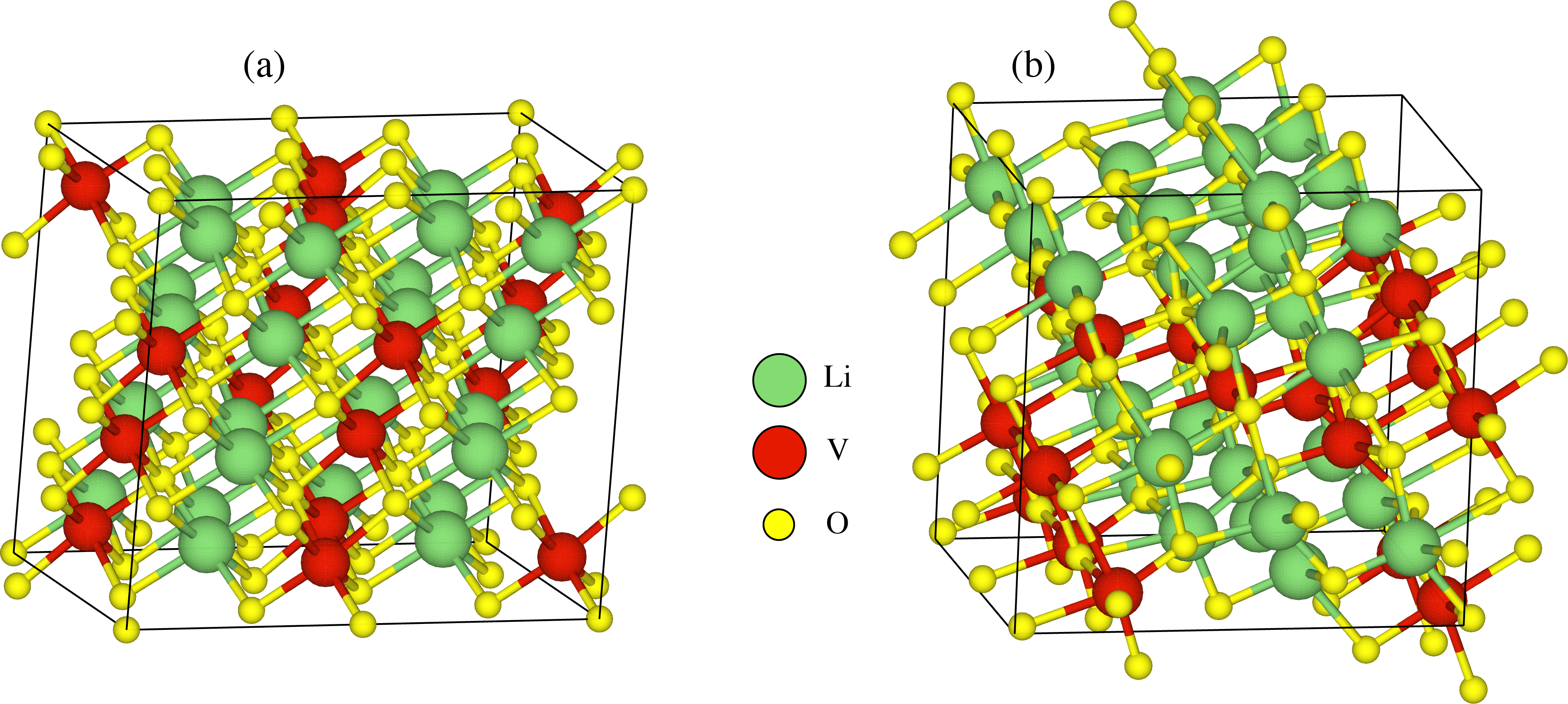}
\caption{\label{fig: system} Supercells of ordered (left) and disordered (right) configurations corresponding to lowest and highest formation energies respectively. Both contain 8 formula units (80 atoms) of \ce{Li3V2O5}, large enough to avoid adjacent image interactions. Ordered system has lithium atoms placed at periodic locations within the supercell, and anionic octahedrons are close to the ideal geometry with mixed Li/V neighbours. Disordered system has irregularly placed lithium with high structural distortion and segregated lithium (\ce{OLi6}) octahedrons.}
\end{figure*}

The extent of cationic disorder in \ce{Li3V2O5} can be organized by its formation energy and bonding character of oxygen atoms. From figure \ref{fig: fE_ox}, we observe that the formation energies vary almost linearly with the average population of electrons in oxygen 2p-orbitals. Since stable configurations have higher electron population, we can deduce that oxygen atoms approach \ce{O^{2-}} state in lower energy disorders due to greater electron sharing by vanadium. Stronger covalent bonding by a mixed cationic environment allows an ideal octahedron geometry around oxygen atoms with almost equal bond lengths and perpendicular angles. In contrast, structures with lesser electrons in the oxygen \textit{p}-shell mimic a peroxo-type state and are characterized by reduced stability and higher octahedral distortion. Therefore, average electron population in anions can be used as a descriptor for stability analysis of disordered configurations. To compare between order and disorder, we have chosen the edge points with the highest and lowest formation energies, respectively, as marked in figure \ref{fig: fE_ox}. Corresponding to these points, supercells containing 8 formula units of \ce{Li3V2O5} (Fig. \ref{fig: system}) were used as reference systems for further calculations. We assume the properties of in-between configurations to be intermediate of that of the edge points. All the configurations contain fully filled octahedral sites (0.6:0.4 ratio of Li:V) and unreserved tetrahedral sites that allow additional lithium insertion. 

Figure~ \ref{fig: fE_ox} shows that the most ordered (leftmost) structure has a considerable energy gap with the rest of the structures, indicating a finite amount of short-range order in this configuration. At room temperature, however, more disorders can have energies close to the ordered structure due to the configurational entropy from cationic rearrangement ($H - TS$) \cite{sarkar2018high}. The dashed line marks the boundary on the left of which structures are thermodynamically accessible from configurational entropy at room temperature. Temperature fluctuations during charge-discharge cycles could further increase disorder resulting in higher formation energies.

Nearest neighbour environments of tetrahedral sites are more uniform in the ordered system allowing minimal segregation into \ce{Li4} (0TM) tetrahedrons since electrostatic attraction of \ce{V^{3.5+}} with \ce{Li^{+}} dominates over the opposing size effect \cite{ji2019hidden}. The mismatch in size between vanadium (0.78 Å \cite{almessiere2019microstructural}) and lithium (0.76 Å \cite{ji2019hidden}) ions is small, allowing lesser strain with mixed Li-V tetrahedrons, thus increasing the energy to segregate. As expected, higher energy disorders have greater concentration of 0TM tetrahedrons.

Octahedral distortion at each anionic site can be quantified by transforming the displacement vectors of the neighbours from their ideal positions into orthogonalized normal coordinates decomposing them toward recognizable modes \cite{urban2017electronic}. Table \ref{tbl: distortion modes} compares the L2 norm of coefficients in different modes for the most distorted octahedron, ordered, disordered and doped systems, respectively. Clearly, the ordered system has much lower magnitude of coefficients than its counterpart across all modes. Higher distortion in disordered systems is responsible for hysteresis and higher chances of irreversible structural transformation \cite{assat2018fundamental}. 

\begin{table}
\centering
\caption{\label{tbl: distortion modes} Normal (symmetrical stretching, asymmetrical stretching, bending, twisting and displacement) mode coefficients of the most distorted anion atom in the ordered, disordered and (two anion and two cation) doped systems. Combined L2 norm of all the modes are listed as column \textit{c} norm.}
\renewcommand{\arraystretch}{1.5}
\begin{tabular}{|l|l|l|l|l|l|l|}
\hline
\textbf{System} & \textbf{\textit{c} norm} & \textbf{Sym. stretch} & \textbf{Asym. stretch} & \textbf{Bend} & \textbf{Twist} & \textbf{Displ.} \\
\hline
Ord. \ce{Li3V2O5}  & 0.547 & 0.0652 & 0.374 & 0.220 & 0.226 & 0.248 \\ 
Dis. \ce{Li3V2O5}   & 1.676 & 0.289 & 0.640 & 0.693 & 0.968 & 0.725 \\
Dis. \ce{Li3V2O4F}(1)  & 2.043 & 0.324 & 0.843 & 0.757 & 1.175 & 0.887 \\
Dis. \ce{Li3V2O4F}(2)  & 0.696 & 0.141 & 0.176 & 0.285 & 0.399 & 0.321 \\
Dis. \ce{Li3TiVO5}  & 1.621 & 0.050 & 0.323 & 0.877 & 0.447 & 0.715 \\
Dis. \ce{Li3NiVO5}  & 2.376 & 0.008 & 0.454 & 0.751 & 0.462 & 0.939 \\

\hline
\end{tabular}
\renewcommand{\arraystretch}{1}
\end{table}

\subsubsection{Electronic structure} \label{ES}

\begin{figure*}[hbt!]
\centering
\includegraphics[width=6.8in]{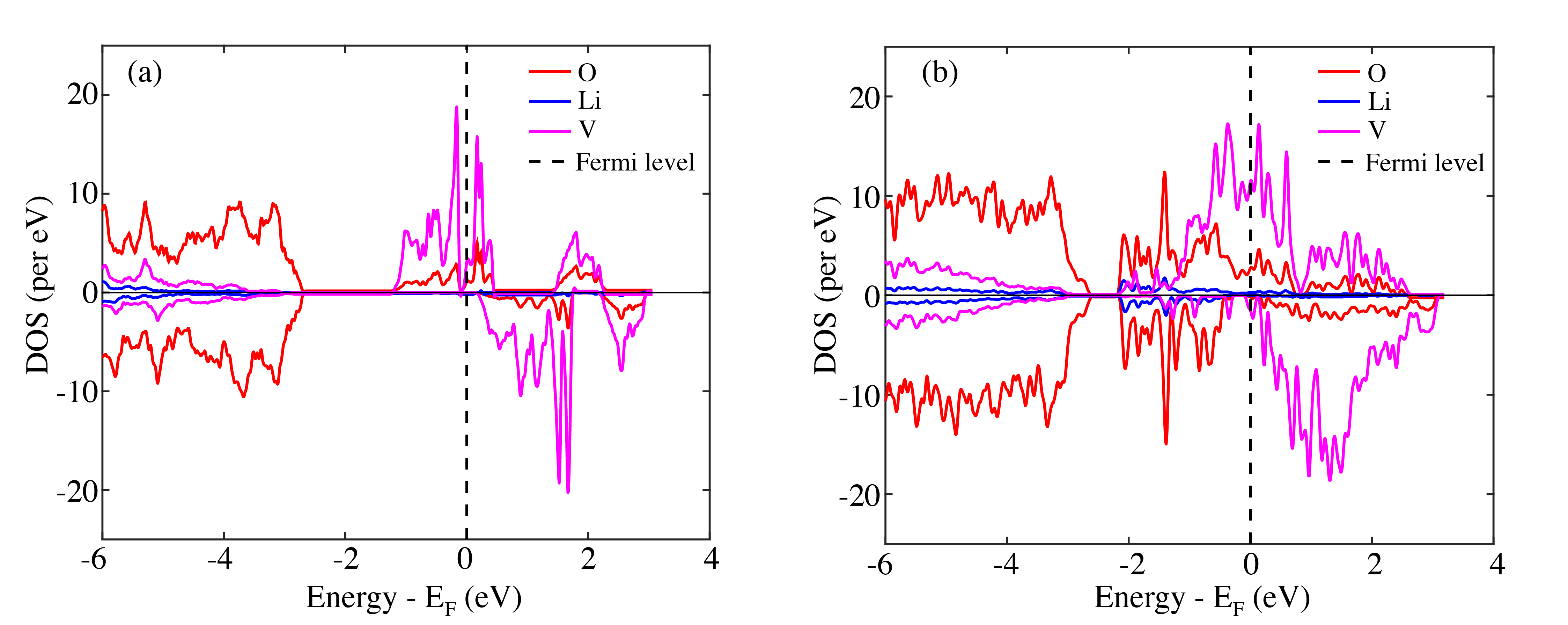}
\caption{\label{fig: dos} Atom projected density-of-states of the ordered (a) and disordered (b) configurations. The disordered structure has mostly filled unhybridized oxygen states (spikes) near the Fermi level ($E_{F}$), caused by oxygen atoms surrounded by lithium only. Mixed Li/V octahedral environment in the ordered system increases covalent sharing with oxygen and delocalizes its states to lower energies. Consequently, average population in oxygen \textit{p}-orbitals is lower in the disordered system since some non-bonding states are unfilled near the Fermi level.}
\end{figure*}

The configuration of the nearest neighbours play an important role in determining the contribution of oxygen in M-O* states in the electronic structure \cite{assat2018fundamental, seo2016structural}. Projected DOS for the ordered and the disordered systems are shown in Fig.~\ref{fig: dos}. Unlike its counterpart, the ordered system seems almost insulating due to lack of states near the Fermi level, which could hinder lithium transport. The disorder shows a metallic behaviour, better for fast charging applications. We shall investigate lithium transport differences more in the migration barriers section.

In the ordered system, we have mixed Li-V environment where none of the oxygen atoms form an \ce{OLi6} octahedron. As expected, we observe delocalized 2p oxygen states in its pDOS (Fig. \ref{fig: dos} (a))  \cite{seo2016structural}. Consequently, we anticipate negligible anionic redox in the ordered configuration due to the absence of redox-active non-bonding states in oxygen atoms.
In contrast, the disordered configuration has a variety of nearest-neighbour environments. Due to lithium segregation, we can identify \ce{OLi6} octahedrons in the unit cell, from which we expect non-bonding 2p oxygen states to arise near the Fermi level. This character is reflected in the pDOS of oxygen (figure \ref{fig: dos} (b)) near the Fermi level. One of the important descriptors of reversible anionic redox is the presence of both transition metal and anion states around the Fermi level for electron extraction during the delithiation process \cite{assat2018fundamental}. More specifically, the non-bonding oxygen 2p orbitals must be partially filled near the Fermi level to facilitate electron addition to these states. As lithium is introduced, these 2p states get filled and shift deeper below the Fermi level, as identified for \ce{Li2RuO3} \cite{saubanere2016intriguing}. However, we find limited participation of these non-bonding oxygen states in the disordered structures where these states are already filled and stay below the Fermi level.

In Figure S1, the crystal orbital overlap population (COOP) of O-O bonds in the pristine disorder is similar in magnitude to the ordered structure. The average oxygen bond length is also shorter ($2.55$ vs $2.69$ Å). To test if the partial O-O bonds in the disorder participate in the redox, we sequentially inserted (and relaxed) two lithium atoms ($x=0.125$) in energetically favourable tetrahedral sites, and observed the O-O bond COOP magnitudes. As evident in figure S2, the overlap populations do not change significantly, which is in contrast with the drastic suppression of O-O COOP in \ce{LiRuO3} vs \ce{RuO3} \cite{saubanere2016intriguing}, where electron fills the empty 2p states, subsequently breaking peroxo O-O bonds on lithium insertion. Based on the pDOS and COOP analysis, we conclude that the unhybridized oxygen states in the disordered system are not destabilized enough to participate in the redox at 0K. However, these states are close enough to get activated at higher temperatures, with voltage fluctuations that shift the Fermi level, and result in surface densification \cite{denis2009electrochemical, assat2018fundamental} during charge-discharge cycles. We expect a similar trend to hold in other disordered configurations lying at intermediate energies (see Fig.~\ref{fig: fE_ox}). 

Another way to engineer activity of redox centers is through cationic and anionic substitution, which we explore next. We shall investigate the effect of doping on the electronic structure in terms of the newly introduced redox-active states as well as the accompanied change in structural distortion. We study the electronic structure change due to anionic substitution by fluorine and cationic substitution by 3d transition metals.  
Fluorine being a more electronegative specie than oxygen, decreases average anionic charge and increases metal redox capacity in TMOs \cite{clement2020cation}.

\textbf{Effect of Fluorination}\\
 
Partial fluorine substitution of oxygen atoms can stabilize or suppress anionic states to allow greater cationic redox and reduce anion participation under thermodynamic activation (high temperature, voltage fluctuations etc.). Previous works have shown high increase in capacity via fluorine substitution of Manganese based TM oxides ($>$ 300 mah g$^{-1}$)\cite{clement2020cation}. With vanadium, the capacity increase has been reported to be higher, from 295 mAh g$^{-1}$ in \ce{Li2VO3} to 420 mAh g$^{-1}$ in \ce{Li2VO2F} \cite{clement2020cation, chen2015disordered}. Presence of vanadium as a high-valent, redox active TM also supports charge compensation on the cation and reduces anionic activity \cite{kitchaev2018design}. To understand this effect, we prepared two fluorine doped structures of the disordered system replacing 20\% oxygen atoms (\ce{Li3V2O4F}), one at random positions, another at selected octahedral sites with majority lithium nearest neighbours (\ce{FLi6} and \ce{FLi5V}), as shown in Fig.~4(a) and 4(b), respectively. 

\begin{figure*}[hbt!]
\centering
\includegraphics[width=6.8in]{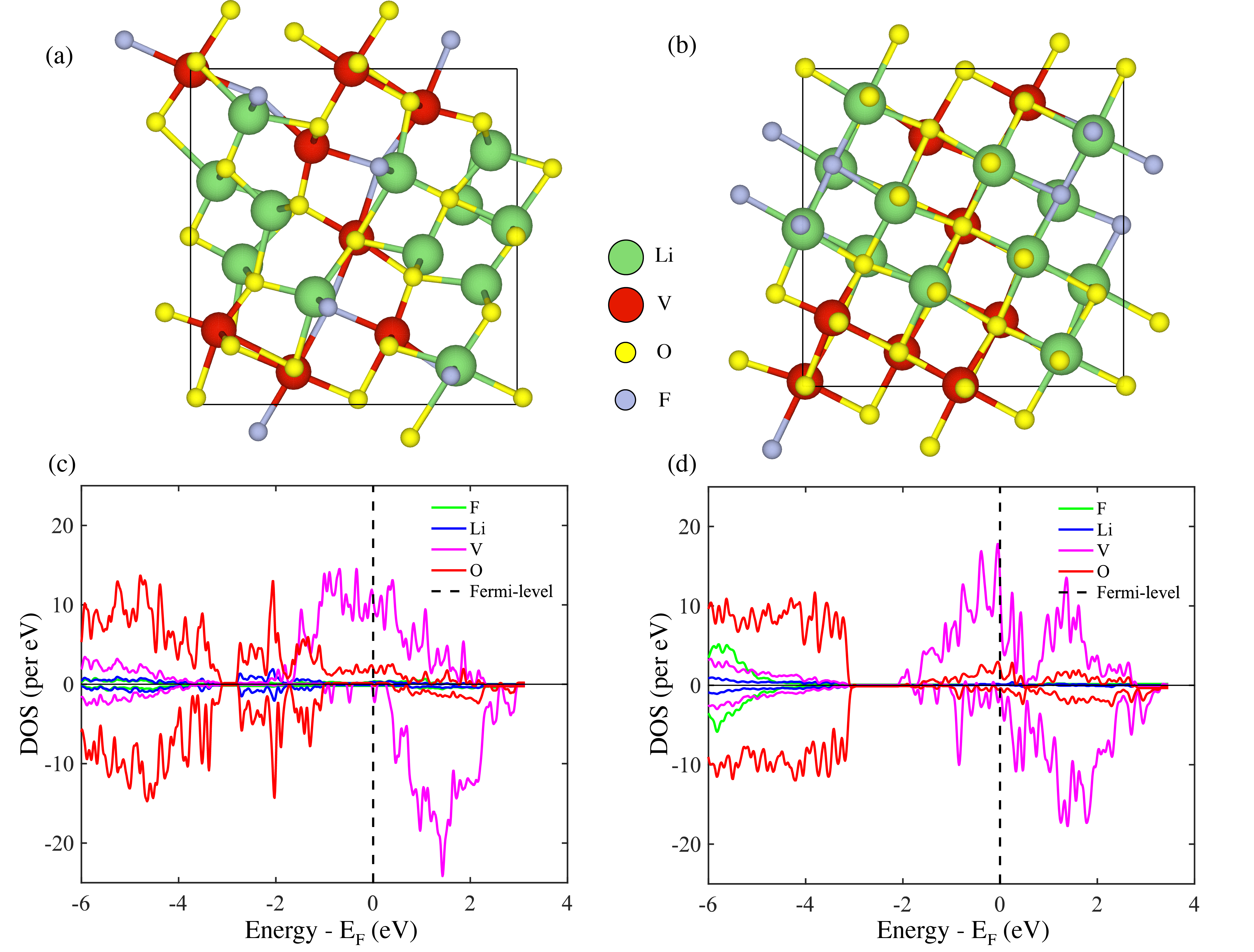}
\caption{\label{fig: f_subst} Atom projected density-of-states for the two cases of fluorinated-disordered \ce{Li3V2O4F} systems. Randomly doped Fluorine (a) maintains octahedral distortion and pushes unhybridized 2p oxygen states deeper into the Fermi level (c). When Fluorine occupies sites with all-lithium nearest neighbours (\ce{FLi6}), distortion is reduced (b) and all oxygen non-bonding states are suppressed near the Fermi level (d). Due to high electron affinity, non-bonding fluorine states occur at much lower energies which are visible as increased DOS magnitude around -6 eV in (d).}
\end{figure*}

Figure \ref{fig: f_subst} shows these configurations in relaxed positions and the corresponding pDOSs. Fig~ 4(c) shows that Fluorine substitution increases the density of active cation states and stabilizes the non-bonding oxygen states by shifting them further below the Fermi level compared to the undoped case (fig. \ref{fig: dos}(b)). On the other hand, a greater suppression of anionic activity is shown by the structure in Fig.~4(b) where the fluorine atoms have eliminated all 2p-oxygen states near the Fermi level, facilitating pure cationic redox over the charge transfer cycle (see pDOS in Fig.~4(d)). This is the result of replacing non-bonding lone pairs of oxygen with the highly electronegative pairs of fluorine, thereby burying these states deep in energy. One can observe an increased DOS of fluorine around $-6$ eV (fig. \ref{fig: f_subst}(d)) from the Fermi level, which are nowhere near active for redox.      

Interestingly, structural distortion increases in the randomly placed fluorine system (Fig.~4(a)) as compared to the undoped case (22\% jump in $c$ norm), even though it stabilizes oxygen lone pairs. This is possibly due to the large electronegativity of fluorine attracting free lone pairs of 2p-oxygen orbitals, producing irregularly mixed octahedrons. This might also be one of the reasons behind severe capacity fade and structural degradation in \ce{Li2VO2F}, assisting rapid O loss from the surface \cite{clement2020cation}. In contrast, when these oxygen atoms with non-bonding electrons are replaced with fluorine (Fig. \ref{fig: f_subst}(b)), the distortion is drastically reduced (59\% and 66\% drop in $c$ norm with respect to undoped and randomly flourinated cases, respectively). All other distortion modes also show a similar drop in coefficients. In essence, deliberate selection of fluorine doping sites can subdue all anionic activity and prevent structural degradation accompanied by disorder. This property is crucial for mitigating hysteresis, ion migrations and capacity fade over cycles, which have been attributed to anionic redox activity in TMOs \cite{assat2018fundamental}.

\begin{figure*}[hbt!]
\centering
\includegraphics[width=6.9in]{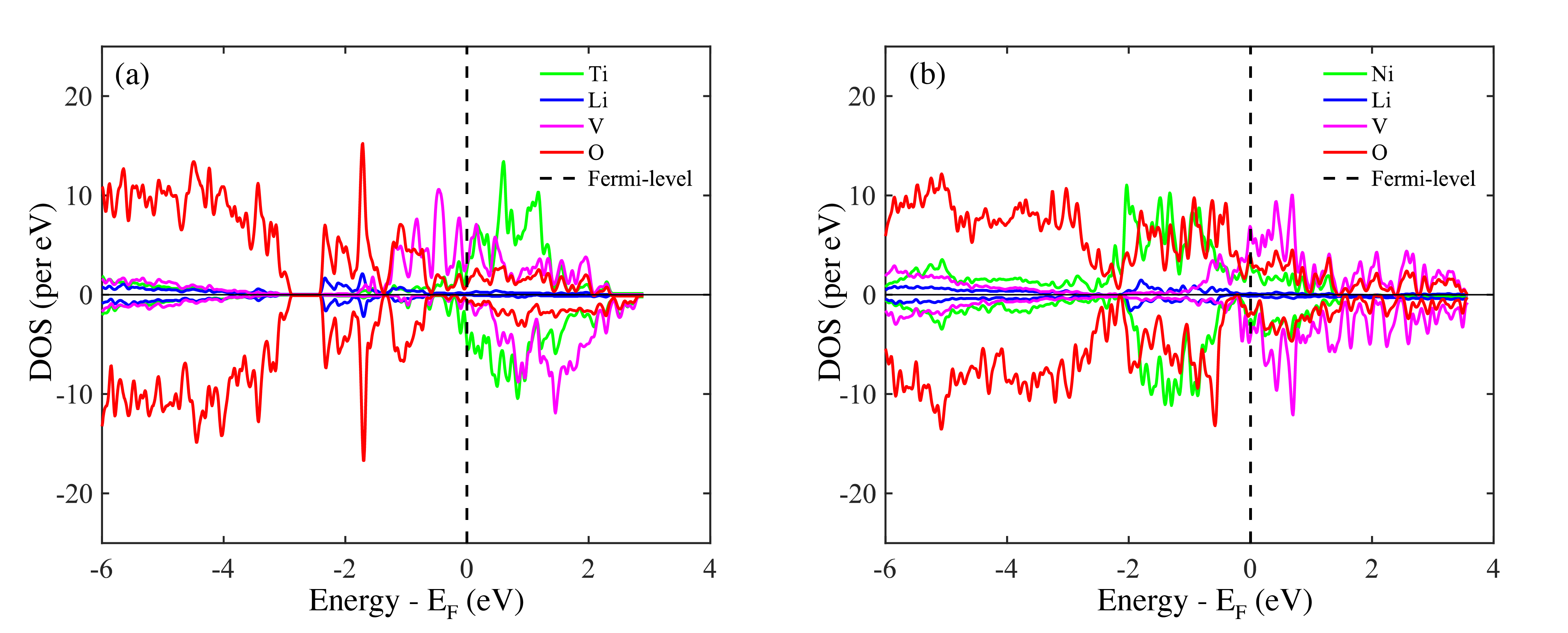}
\caption{\label{fig: ti_ni} Effect of 50\% cationic doping Titanium (left) and Nickel (right) on atom projected DOS of disordered \ce{Li3VTMO5}. Titanium stabilizes oxygen non-bonding electrons and increases cation states in the conduction band, assisting cationic redox. In contrast, most Nickel states are already filled and non-bonding oxygen states are closer to the Fermi level, supporting anionic redox.}
\end{figure*}

\textbf{Transition metal substitution}\\
Transition metal substitution can help to facilitate or diminish anionic redox by either activating or stabilizing oxygen states near the Fermi level. Alternatively, the participation of the TM states in redox process depends on the configuration of the 3d states outer shell.
For example, the ratio of cationic vs anionic DOS above the Fermi level is lower in Ni as compared to Ti since they are mostly filled and inactive, as shown by 50\% doped \ce{Li3VTMO5} in Fig.~\ref{fig: ti_ni}. Ni also pushes unhybridized oxygen states closer to the Fermi level, even though none of the 3d metals activate the non-bonding 2p states above the Fermi level (figure S4). 

Similar to fluorine substitution at selected sites, structural distortion decreases with Ti doping due to the stabilization of oxygen states, though it is small (4\% drop in overall $c$ norm as compared to the undoped case, table \ref{tbl: distortion modes}) due to higher mismatch in size between \ce{Ti^{4+}} (0.605 Å) and \ce{Li^{+}} (0.76 Å) \cite{ji2019hidden}. On the other hand, both factors (anionic activation and size effect) add together to increase distortion from Ni doping, producing a huge jump of 40\% in $c$ norm. In such systems, we expect an increased tendency of lithium to segregate as \ce{Li4} tetrahedrons to maintain homogeneity of nearest neighbours. Further studies are required to assess behaviour in other metrics like OCV, volume and stresses changes with TM doping.

\subsubsection{Migration Barriers} \label{section:mb}

\begin{figure*}[hbt!]
\centering
\includegraphics[width=6.9in]{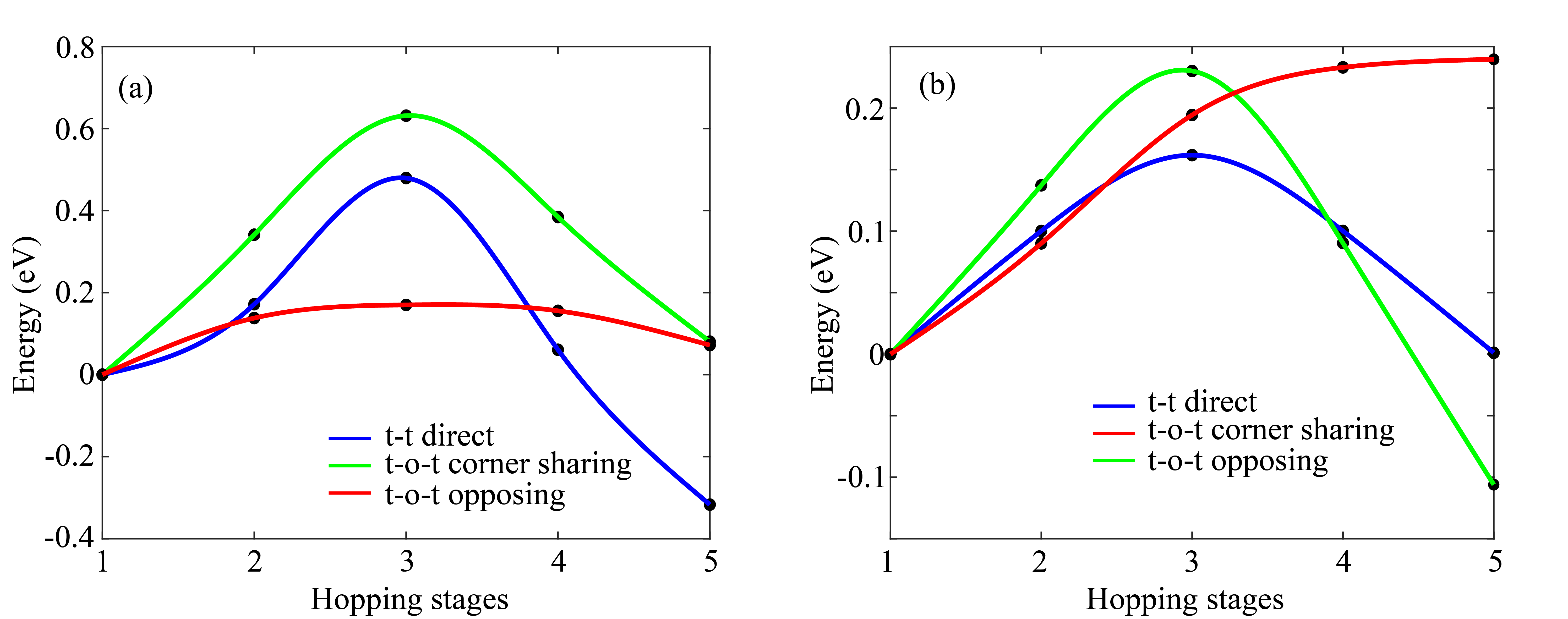}
\caption{\label{fig: neb} NEB migration barriers of the ordered (left) and disordered (right) configurations for the three lithium transition paths, direct t-t, obtuse t-o-t and opposing t-o-t. Only one path (opposing t-o-t) is feasible in the ordered system, while all paths become available in the disorder due to a drastic drop in activation energy caused by weak covalent bonds. Opposing t-o-t can be spontaneously driven (0 meV). Barriers for the other two paths (150 meV and 230 meV resp.) fall in the feasible range, being much smaller than that of graphite (420 meV).}
\end{figure*}

Order-to-disorder transition is also accompanied by a change in diffusion properties. Macroscopic diffusion is more readily satisfied in Li-excess materials with long distance connectivity of segregated 0TM sites \cite{urban2014configurational, ji2019hidden}. \ce{Li3V2O5} has 20\% excess lithium ($x = 1.2$ in \ce{Li_{x}V_{2-x}O2}) which is higher than the threshold (9\%) identified by \citet{urban2014configurational} for activating long distance percolation in DRS assuming only 0TM lithium diffusion. Microscopic diffusion can be better quantified with migration barrier calculations for lithium transition from one stable site to another. Results of NEB calculations on pristine (\ce{Li3V2O5}) phases for two different configurations (most stable and least stable disorders) are shown in Fig.~\ref{fig: neb}. We have compared the three known lithium transition paths for the two systems; direct t-t, corner sharing t-o-t and opposing t-o-t. Confirming previous studies \cite{liu2020disordered}, t-o-t opposing knock off is the most favourable path ($\sim$ 150 meV) in ordered \ce{Li3V2O5} due to least steric hindrance and Coulombic repulsion by vanadium atoms. Direct t-t and corner sharing t-o-t are at least 300 meV higher in activation energy. In the disordered system, however, all paths dramatically shift to lower energies; opposing t-o-t undergoes almost spontaneous transition (0 meV), while direct t-t (150 meV) and corner sharing t-o-t (230 meV) paths are almost 50\% lesser than lithium diffusion through Frenkel mechanism in graphite (420 meV) \cite{thinius2014theoretical}). This should also shift the mass transport limit higher than graphite, providing more resistance against lithium plating \cite{petzl2015lithium}.  
The crystal orbital Hamiltonian overlap (COHP) of summed V-O bonds in the disordered structure is also significantly smaller as compared to its counterpart (fig. S3). This indicates higher structural flexibility in the disordered system. Hence, higher distortion and weaker TM-O bonds facilitate easier lithium diffusion in the originally hindered migration paths. Consequently, all three paths get activated in the disordered system, leading to higher diffusion rates and faster charging ability.

\subsection{Conclusions}
We have quantified structural distortion, analyzed electronic structure, and microscopic diffusion barriers in ordered and disordered \ce{Li3V2O5}, which is the delithiated starting point in the $\omega$-phase. The disorder is characterized by segregated lithium, weaker TM-O covalent bonds, disorganized octahedrons and potentially redox-active oxygen states. In contrast, the ordered structure possesses mixed Li-V octahedrons with similar ionic radii, producing highly covalent TM-O bonds matching ideal octahedron geometry. The oxygen orbitals are delocalized and approach $O^{2-}$ state, as evidenced by the higher average electron population in its oxygen p-orbitals. Using pDOS and COOP analysis we can identify the positions of unhybridized oxygen 2p states and the associated changes in O-O bond strengths. At 0K, these states are chemically inactive under high-energy disorder, which can get activated with voltage/temperature fluctuations. Additionally, we have found a positive trend between structural distortion and activation of anionic states using our first-principles analysis. This correlation is supported by anionic (fluorination) and cationic (3d metals) doping whereby replacement of selected oxygen sites significantly decreases distortion and anionic activity. We expect this suppression of anionic activity will play an important role in reducing hysteresis and voltage fade in high energy disorders \cite{assat2018fundamental}. In terms of lithium transport, higher energy disorders show better promise due to their metallic character and structural flexibility. Lower TM-O bond strength and hindrance has reduced the activation energies of all 3D-lithium hoppings, essential for fast charging \cite{liu2020disordered}. Hence we encounter two effects from disorder-induced structural distortion in \ce{Li3V2O5}; it activates anionic states for potential redox and improves lithium transport. 

It will be interesting to understand the effect of cationic and anionic doping on other performance metrics like OCV, volume and stress changes, migration barriers etc., although these computations entail high computational cost. The problem could, however, be addressed by using state of the art machine-learning architectures as an alternative to DFT for high throughput calculations. With novel Monte Carlo techniques we can identify stable lithium intercalation sites and variation in disorder as a function of the charge-state and report on-the-fly convex hull of free energies and OCV \cite{babar2020accurate}. One can also investigate other dopants like 4d and 5d metals, and anions other than fluorine. Since layered \ce{V2O5} is a popular cathode, an all-vanadate based lithium-ion battery can be an important electrode combination to explore. Overall, our analysis should be relevant for experimentalists and theorists to further investigate the potential of doped disordered-\ce{Li3V2O5} as durable, fast charging anodes. Due to low operational voltage (0.4 V), the present system with appropriate modifications is ideal for low temperature and fast charging applications, since it provides thermodynamic safety at weak kinetic conditions and faster lithium transport than graphite. 


\begin{acknowledgement}
We thank Dr. Zhuoying Zhu and Dr. Gus Hart for helpful discussions. M. B. and V. V. acknowledge support from the Office of Naval Research under Award No. N00014-19-1-2172. Acknowledgment is also made to the Extreme Science and Engineering Discovery Environment (XSEDE) for providing computational resources through Award No. TG-CTS180061. B. B. is supported by Ministry of Education and Culture (Finland). A.B. acknowledges support by the US Department of Energy (DOE), Office of Science, Basic Energy Sciences grant number DE-FG02-07ER46352, and Northeastern University's Advanced Scientific Computation Center (ASCC) and the NERSC supercomputing center through DOE grant number DE-AC02-05CH11231.

\end{acknowledgement}

\begin{suppinfo}
Supporting Information (suppinfo.pdf) contains the following details.
\begin{itemize}
\item Summed COOP and COHP between O-O/V-O bonds in the disordered structure. 
\item Electronic DOS for 3d-metal substituted disordered system other than Ti and Ni. 
\item Filled out checklist.\cite{Mistry2021}
\end{itemize}

\end{suppinfo}


\bibliography{cite}

\end{document}


\beginsupplement

\clearpage


We used the LOBSTER code to obtain the summed overlap population of O-O/V-O bonds in the system.

\begin{figure*}[hbt!]
\centering
\includegraphics[width=4.5in]{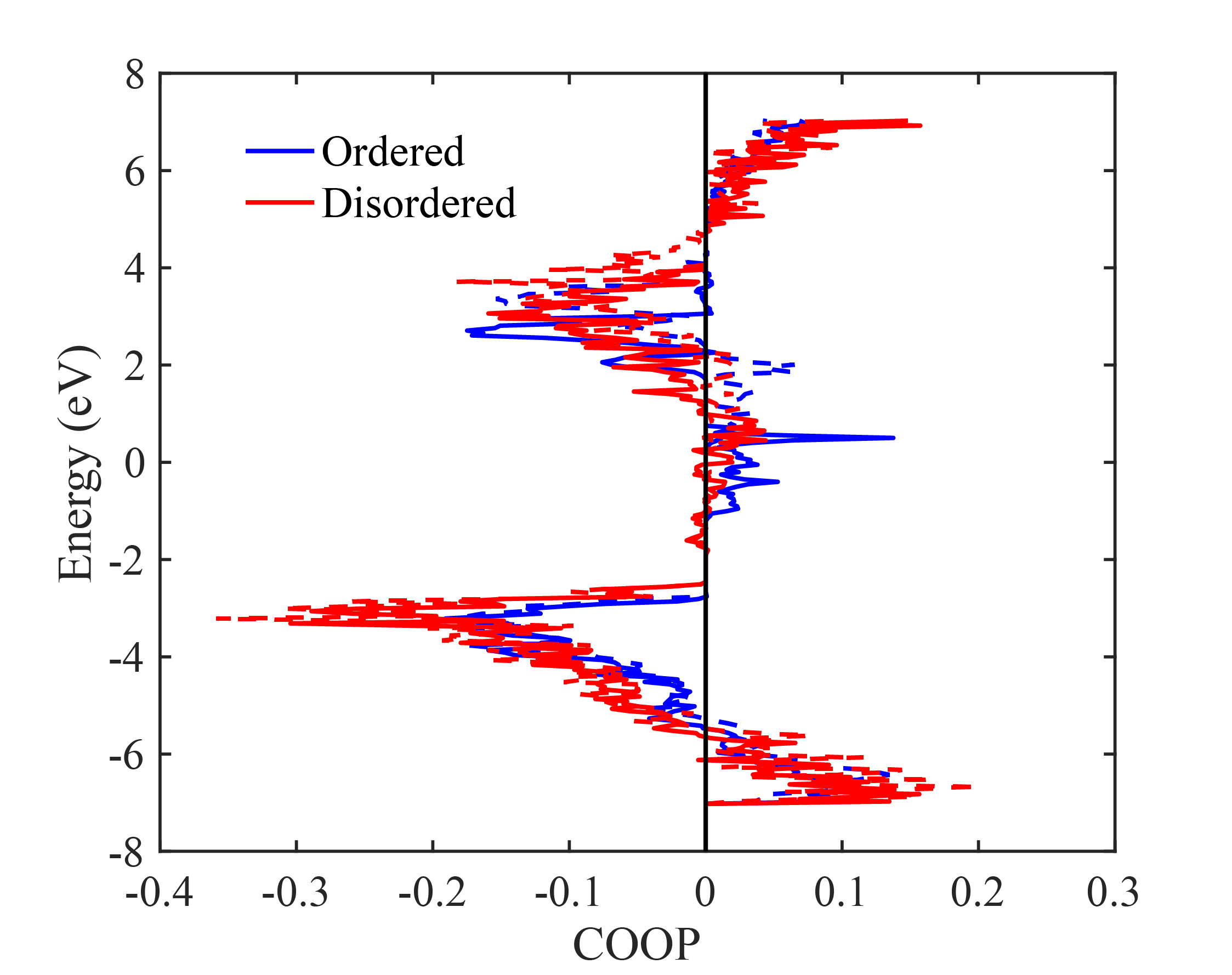}
\caption{\label{fig: s1} Summed crystal orbital overlap population of O-O bonds in most ordered (blue) and disordered (red) configurations of \ce{Li3V2O5}. Negative and positive overlap mark the bonding and anti-bonding regions respectively. The magnitudes of overlap are similar in the two cases, indicating minimal differences in oxygen bonding.}
\end{figure*}

\begin{figure*}[hbt!]
\centering
\includegraphics[width=7.0in]{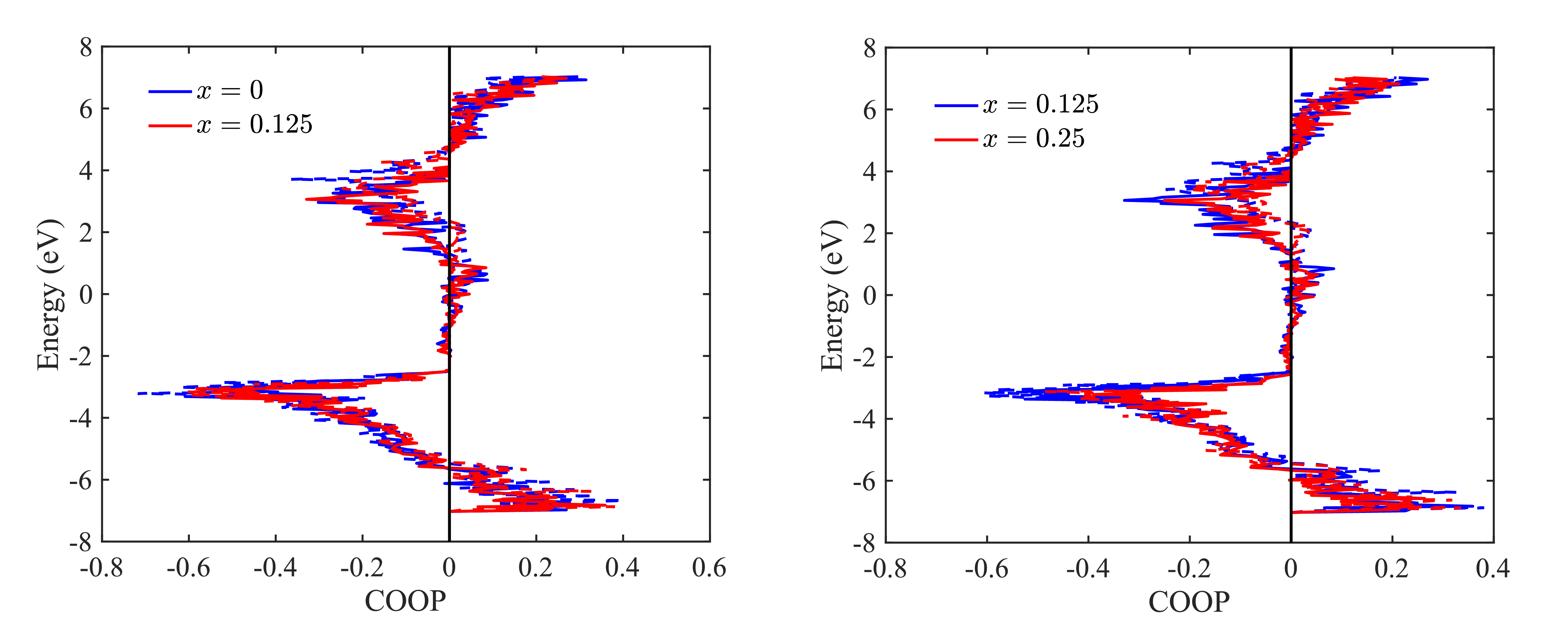}
\caption{\label{fig: s2} Comparison of two successive lithium addition on summed COOP of O-O bonds in disordered \ce{Li3V2O5}. There is a negligible change in COOP magnitudes, supporting predominant cationic redox in the early part of lithiation.}
\end{figure*}
 
\begin{figure*}[hbt!]
\centering
\includegraphics[width=4.5in]{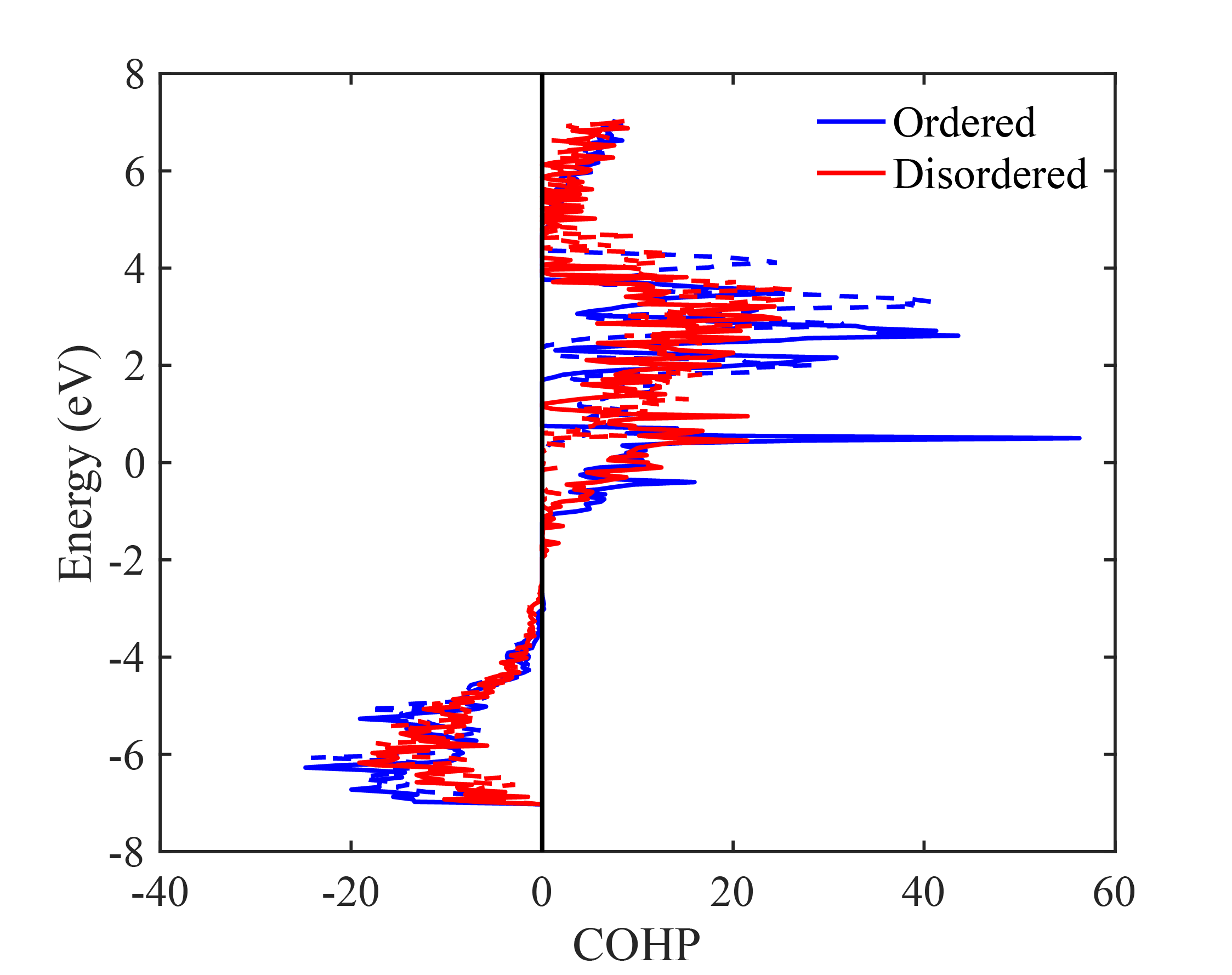}
\caption{\label{fig: s3} Summed crystal hamiltonian overlap population of V-O bonds in most ordered (red) and disordered (blue) configurations of \ce{Li3V2O5} respectively. The magnitude of V-O overlap in the disordered system is smaller than its counterpart, highlighting weaker TM-O bonds due to structural distortion.}
\end{figure*}

\newpage

Projected DOS of the disordered system doped with a 3d metal (\ce{Li3VTMO5}) other than Ti/Ni (in the main text).

\begin{figure*}[hbt!]
\centering
\includegraphics[width=6.8in]{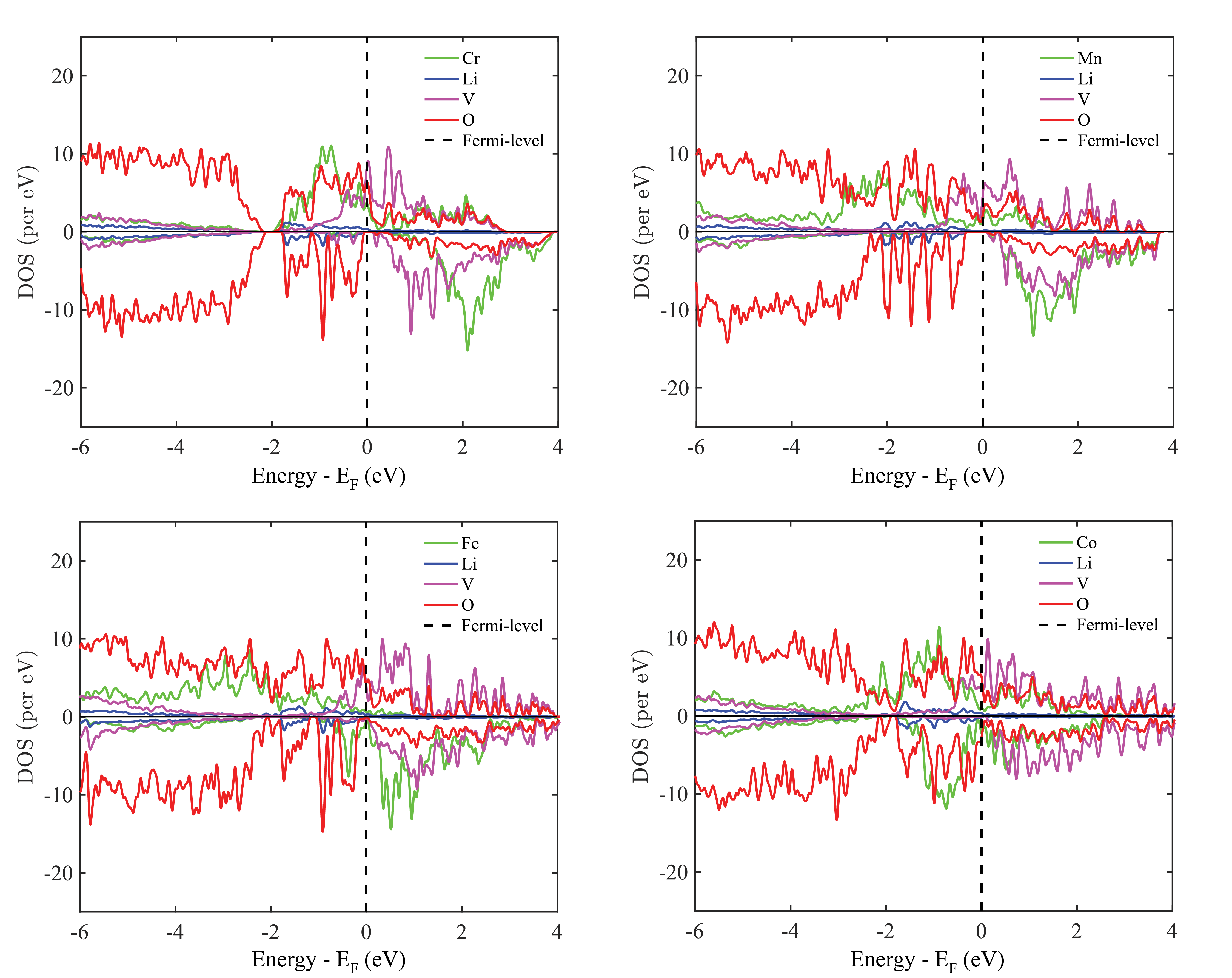}
\caption{\label{fig: s4} Projected density of states of the disordered system (\ce{Li3VTMO5}) 50\% doped with 3d metals (Cr, Mn, Fe, Co) in order.}
\end{figure*}

\newpage
\documentclass[SM.tex]{subfiles}

\renewcommand{\arraystretch}{1.5}

\begin{document}

\begin{table}[!h]
    \begin{tabular}{@{}p{2in}p{4.5in}@{}}
        \toprule
        Manuscript Title: &
        Effect of disorder and doping on electronic structure and diffusion properties of \ce{Li3V2O5} 
        \\ \midrule
        Submitting Author${}^{\ast}$: & 
        Mohammad Babar
        \\ 
    \end{tabular}

    \begin{tabular}{@{}lp{5.5in}c@{}}
        \toprule
        \# & Question & Y/N/NA${}^{\dagger}$ \\ \midrule
        1 & Have you provided all assumptions, theory, governing equations, initial and boundary conditions, material properties, e.g., open circuit potential (with appropriate precision and literature sources), constant states, e.g., temperature, etc.? & 
        Y
        \\
        & \multicolumn{2}{p{6.3in}}{\textbf{Remarks:}All major assumptions, theory and equations are provided in the main text. Material properties like OCV, volume and stress changes as a function of lithium in DRS-\ce{Li3V2O5} have been reported in previous works.}  \\

        \midrule
        2 & If the calculations have a probabilistic component (e.g. Monte Carlo, initial configuration in Molecular Dynamics, etc.), did you provide statistics (mean, standard deviation, confidence interval, etc.) from multiple $\left( \geq 3 \right)$ runs of a representative case? & 
        NA
        \\
        & \multicolumn{2}{p{6.3in}}{\textbf{Remarks:} } \\

        \midrule
        3 & If data-driven calculations are performed (e.g. Machine Learning), did you specify dataset origin, the rationale behind choosing it, what all information does it contain and the specific portion of it being utilized? Have you described the thought process for choosing a specific modeling paradigm?  & 
        NA
        \\
        & \multicolumn{2}{p{6.3in}}{\textbf{Remarks:} } \\

        \midrule
        4 & Have you discussed all sources of potential uncertainty, variability, and errors in the modeling results and their impact on quantitative results and qualitative trends? Have you discussed the sensitivity of modeling (and numerical) inputs such as material properties, time step, domain size, neural network architecture, etc. where they are variable or uncertain? &
        Y
        \\
        & \multicolumn{2}{p{6.3in}}{\textbf{Remarks:} Variability in the findings can come from using different exchange correlation functionals in the DFT calculations of energies, forces and electronic structures. For this, we have provided sufficient evidence in support of GGA+\textit{U} functional in the methods section. } \\

        \midrule
        5 & Have you sufficiently discussed new or not widely familiar terminology and descriptors for clarity? Did you use these terms in their appropriate context to avoid misinterpretation? Enumerate these terms in the `Remarks'. & 
        Y
        \\
        & \multicolumn{2}{p{6.3in}}{\textbf{Remarks:} All terminologies are derived from previous works.} \\ \bottomrule
    \end{tabular}
    
    {
    \footnotesize
    ${}^{\ast}$ I verify that this form is completed accurately in agreement with all co-authors, to the best of my knowledge.
    }
    
    {
    \footnotesize
    ${}^{\dagger}$ Y $\equiv$ the question is answered completely. Discuss any N or NA response in `Remarks'.
    }
\end{table}

\end{document}
